\documentclass[a4paper]{jpconf}

\usepackage{braket,bm,subfig}

\usepackage{tikz}

\usepackage[pass]{geometry}
\usepackage{color}
\usepackage{mathrsfs,mathtools, wasysym,amsmath, amsfonts,amssymb,graphicx}
  \DeclareMathAlphabet{\mathpzc}{OT1}{pzc}{m}{it}

 \usepackage{textpos}
\makeatletter

\def\ps@headings{%
     \def\@oddfoot{\hfil\thepage\hfil}
     \def\@evenfoot{\hfil\thepage\hfil}
     \let\@oddhead\@empty
     \let\@evenhead\@empty
      \let\@mkboth\markboth
      \let\sectionmark\@gobble
      \let\subsectionmark\@gobble}

\makeatother

\newcommand{\CP}{C\!P}
\newcommand{\maf}[1]{\mathbf{#1}}

\begin{document}



\title{TeV Scale Lepton Number Violation and Baryogenesis}

\author[a]{P.~S.~Bhupal Dev$^1$, Chang-Hun Lee$^2$, \underline{R. N. Mohapatra}$^2$}

\address{$^1$ Consortium for Fundamental Physics,
  School of Physics and Astronomy, University of Manchester, Manchester M13 9PL, United Kingdom.}

\address{$^2$ Maryland Center for Fundamental Physics and Department of Physics,
University of Maryland, College Park, Maryland 20742, USA.}

\begin{abstract} 
Contrary to the common lore based on naive dimensional analysis, the seesaw scale for neutrino masses can be {\em naturally} in the TeV range,  with small parameters coming from radiative corrections. We present one such class of type-I seesaw models, based on the left-right gauge group $SU(2)_L\times SU(2)_R\times U(1)_{B-L}$ realized at the TeV scale, which fits the observed neutrino oscillation parameters as well as other low energy constraints. We discuss how the small parameters of this scenario can arise naturally from one loop effects. The neutrino fits in this model use quasi-degenerate heavy Majorana neutrinos, as also required to 
 explain the matter-antimatter asymmetry in our Universe via resonant leptogenesis mechanism. We discuss the constraints implied by the dynamics of this mechanism on the mass of the right-handed gauge boson in this class of models with enhanced neutrino Yukawa couplings compared to the canonical seesaw model and find a lower bound of $m_{W_R}\geq 9.9$ TeV for successful leptogenesis assuming maximal $\CP$ asymmetry for each flavor. We also present a model with explicit neutrino mass fit, where the lower bound goes up to 13.1 TeV due to less than maximal primordial $\CP$ asymmetry predicted by the model. 
\end{abstract}


\section{Introduction}
The seesaw mechanism~\cite{seesaw} seems to provide a simple way to understand the smallness of neutrino masses. The simplest among the various seesaw mechanisms is the type-I seesaw, which requires the introduction of Standard Model (SM) gauge-singlet right-handed (RH) neutrinos ($N$'s). The Majorana mass $M_N$ of these RH neutrinos explicitly breaks the global $B-L$ symmetry of the SM by two units and gives rise to a light neutrino mass matrix of the form  
\begin{align}
M_\nu \ \simeq \ -M_D M^{-1}_N M_D^{\sf T} \; , 
\label{Mnu}
\end{align}
where $M_D$ is the Dirac mass generated due to the neutrino Yukawa couplings $Y_\nu$ after the electroweak symmetry breaking. If one assumes maximal allowed values for the third-generation Dirac mass $M_{D,33}$, as is implied by certain Grand Unified Theories such as $SO(10)$, then the seesaw scale is pushed up to $\sim 10^{14}$ GeV making it virtually inaccessible to both collider experiments at the energy frontier and other low energy searches at the intensity frontier. Such naive (``dimensional counting") considerations however can be misleading in the context of specific ultraviolet (UV) complete seesaw models where there may be dynamical suppressions of neutrino Dirac masses. One class of such models that are based on the SM gauge group generally go by the name of radiative seesaw models; see e.g.~\cite{radiative}. 
Within the type-I seesaw framework, there exists another possibility to suppress the light neutrino masses by assigning specific textures to the Dirac and Majorana mass matrices in the seesaw formula~\cite{kersten, texture}. The stability of these textures can in principle be guaranteed by enforcing some symmetries in the lepton sector.  

Extending the SM gauge group to the Left-Right (L-R) Symmetric gauge group  $SU(2)_L\times SU(2)_R\times U(1)_{B-L}$ provides a simple UV-complete seesaw model~\cite{LR}. Here, the RH neutrinos are a necessary part of the model and do not have to be added `by hand' just to implement the seesaw mechanism. An important point is that the RH neutrinos acquire a Majorana mass as soon as the $SU(2)_R$ symmetry is broken at a scale $v_R$. This is quite analogous to the way the charged fermions get masses in the SM by the Higgs mechanism when the $SU(2)_L$ gauge symmetry is broken at a scale $v$. The Higgs field that gives mass to the RH neutrinos becomes the analogue of the 125 GeV Higgs boson discovered at the LHC.

In this proceedings, we focus on a new class of L-R seesaw models~\cite{DLM}, where in addition to the RH gauge bosons as well as other Higgs fields, light (TeV scale) RH neutrinos with sizable mixing with the left-handed (LH) neutrinos may be accessible at colliders and other low energy experiments (for a review, see e.g.~\cite{Deppisch:2015qwa}). With the LHC ramping up to a higher center of mass energy $\sqrt s=13$ TeV and later to 14 TeV, our model as well as other low scale seesaw models, should provide motivation for testing the origin of neutrino mass in the coming years. For the collider phenomenology of TeV-scale LRSM, see e.g.~\cite{KS, LRcol, CDM}, and for the current experimental status, see~\cite{CMS}. Such considerations should also be of interest~\cite{Rizzo:2014xma} for setting physics goals for future higher energy colliders such as the $\sqrt s=80$ or 100 TeV VLHC~\cite{Barletta:2014vea}. It is also important to mention here that the new class of L-R seesaw models with relatively large  left-right neutrino mixing can be distinguished from the minimal LRSM using different kinematic variables~\cite{CDM, Han}. 

An attractive consequence of the seesaw mechanism is the possibility that it could solve another important puzzle of cosmology,~i.e.~the origin of matter-antimatter asymmetry in our Universe, via a mechanism called leptogenesis~\cite{lepto}. At the heart of this mechanism is the out-of-equilibrium decay of the RH Majorana neutrinos via the decay modes $N\to L\Phi, L^{c}\Phi^c$ (where the superscript $c$ denotes the $\CP$-conjugate), which violate $L$, $C$ and $\CP$, thereby dynamically generating an asymmetry in the lepton sector. This primordial lepton asymmetry undergoes thermodynamic evolution as the Universe expands and finally gets converted to a baryon asymmetry via equilibrated electroweak sphaleron interactions~\cite{kuzmin:1985mm}. The attractive aspect of this mechanism is that the {\em same} Yukawa couplings $Y_\nu$ that give rise to neutrino masses via the seesaw mechanism in Eq.~\eqref{Mnu} are also responsible for the origin of matter, thus implying an intimate connection between two seemingly disparate pieces of evidence for beyond the SM physics. This beautiful idea can be tested, provided the seesaw scale, and hence, the scale at which leptogenesis takes place, is accessible to current and near future laboratory experiments. Low-scale leptogenesis is possible when at least two RH Majorana neutrinos have a small mass difference comparable to their decay widths, thus resonantly enhancing the $\varepsilon$-type $\CP$  asymmetry~\cite{Pilaftsis:1997jf, Pilaftsis:2003gt} due to RH neutrino self-energy effects~\cite{Flanz:1994yx}. 

An important question to explore is whether it is possible to falsify the leptogenesis mechanism by laboratory experiments~\cite{Deppisch:2013jxa}. In particular, it is worth investigating if the leptogenesis mechanism works in a TeV-scale realization of L-R seesaw and whether an observation of a RH gauge boson at the LHC could falsify this scenario. This problem was analyzed in detail in~\cite{hambye} and it was pointed out that the additional dilution and washout effects on the lepton asymmetry due to $W_R$-mediated $\Delta L=1$ scattering processes become important for a TeV-scale L-R seesaw. Since in generic versions of the seesaw model, sub-eV neutrino masses would require the Dirac Yukawa couplings to be $Y_\nu \lesssim 10^{-11/2}$,
the washout effects would lead to an efficiency factor $\kappa \sim \frac{Y^2_\nu M^4_{W_R}}{g_R^4 M^4_N}$ ($g_R$ being the $SU(2)_R$ gauge coupling, which is assumed to be equal to the $SU(2)_L$ gauge  coupling in the minimal scenario) which is too small 
for a low-scale leptogenesis to work.  
Through a detailed analysis of the relevant Boltzmann equations,  it was concluded in~\cite{hambye} that even for maximal $\CP$-asymmetry $\varepsilon\sim {\cal O}(1)$, the observed 
value of the baryon asymmetry can be explained by leptogenesis in these L-R models, only if $m_{W_R}\geq 18 $ TeV. Turning this argument around, if a positive signal for $W_R$ is observed at the LHC, this will falsify leptogenesis as a mechanism for understanding the origin of matter in L-R seesaw framework. Since this is such an important issue, we have reanalyzed this~\cite{DLM2} to examine the robustness of the lower bound on $m_{W_R}$ and to investigate if there exists {\em any} allowed parameter space with successful leptogenesis for smaller values of $m_{W_R}$. We find that for the generic class of L-R seesaw models with large light-heavy neutrino mixing, successful leptogenesis requires $m_{W_R}\geq 9.9$ TeV for the maximal $\CP$ asymmetry. For illustration, we will present an explicit neutrino mass fit which satisfies the observed baryon asymmetry. 
The main new results presented here should supersede those reported in~\cite{DLM2}.

 This proceedings is organized as follows: in Section~\ref{sec2}, we review the basic features of the generic L-R seesaw models as well as our new L-R seesaw model with special Dirac and Majorana textures resulting in light neutrinos via type-I seesaw with large light-heavy neutrino mixing generated in a natural manner. In Section~\ref{sec3}, we present an alternative neutrino mass fit relying on cancellations in the seesaw formula. In Section~\ref{sec4}, we discuss resonant leptogenesis in this model and the impact of a TeV-scale $W_R$ on the lepton asymmetry. Our conclusions are given in Section~\ref{sec5}.

\section{A TeV-scale left-right seesaw model}\label{sec2}
In the minimal L-R symmetric model (LRSM), the fermions are assigned to the gauge group $SU(2)_L\times SU(2)_R\times U(1)_{B-L}$ as follows: denoting
$Q_i\equiv (u,d)_i^{\sf T}$ and $\psi_i \equiv (\nu_l, l)_i^{\sf T}$ as the quark and lepton doublets of the $i$th generation respectively, $Q_{L,i}$ and $\psi_{L,i}$ (also denoted simply by $L_i$) are assigned to doublets under the $SU(2)_L$ group, while $Q_{R,i}$ and $\psi_{R,i}$ (also denoted by $R_i$) as the doublets under the $SU(2)_R$ group. 
The Higgs sector of the model can consist of one or several bidoublets $\phi_a$ and triplets $\Delta_{R,b}$:
\begin{align}
\phi \ \equiv \ \left(\begin{array}{cc}\phi^0_1 & \phi^+_2\\\phi^-_1 & \phi^0_2\end{array}\right),
\qquad \qquad
 \Delta_R \ \equiv \ \left(\begin{array}{cc}\Delta^+_R/\sqrt{2} & \Delta^{++}_R\\\Delta^0_R & -\Delta^+_R/\sqrt{2}\end{array}\right).   \label{LRhiggs}
\end{align}
The gauge symmetry $SU(2)_R\times U(1)_{B-L}$ is broken by the vacuum expectation value (VEV) $\langle \Delta^0_R\rangle = v_R$ to the group $U(1)_Y$ of the SM. The LH counterpart ($\Delta_L$) to $\Delta_R$ is not considered here, assuming that parity and $SU(2)_R$ gauge symmetry scales are decoupled. In this case, the $\Delta_L$ fields become heavy when the discrete parity symmetry is broken, and disappear from the low energy theory~\cite{CMP}. 
The VEV of the $\phi$ field given by $\langle\phi\rangle={\rm diag}(\kappa, \kappa')$ breaks the SM gauge group to $U(1)_{\rm em}$. The fermion masses are obtained from the following Yukawa Lagrangian: 
\begin{align}
-{\cal L}_Y \ = \ & h^{q,a}_{ij}\bar{Q}_{Li}\phi_aQ_{R,j}+\tilde{h}^{q,a}_{ij}\bar{Q}_{L,i}\tilde{\phi}_aQ_{R,j}+
h^{l,a}_{ij}\bar{L}_i\phi_aR_j \nonumber \\ 
& \quad + \tilde{h}^{l,a}_{ij}\bar{L}_i\tilde{\phi}_aR_j
+f_{ij} (R_i^{\sf T}Ci\tau_2 \Delta_R R_j +L_i^{\sf T}Ci\tau_2 \Delta_L L_j)+{\rm H.c.},
\label{yuk}
\end{align}
where $\tilde{\phi}=\tau_2\phi^*\tau_2$ ($\tau_2$ being the second Pauli matrix). After electroweak symmetry breaking, the Dirac fermion masses are given by the generic formula $M_f~=~h^f\kappa + \tilde{h}^f\kappa'$ for up-type fermions, and for down-type quarks and charged leptons, it is the same formula with $\kappa\leftrightarrow \kappa'$.  The Yukawa Lagrangian (\ref{yuk}) leads to the Dirac neutrino mass matrix $M_D = h^{l}\kappa + \tilde{h^{l}}\kappa'$ and the Majorana mass matrix for the heavy RH  neutrinos $M_N=fv_R$ which go into the seesaw formula \eqref{Mnu} for calculating the neutrino masses and the heavy-light neutrino mixing. 

Since the Higgs sector relates the neutrino Yukawa couplings with the charged-lepton ones in the LRSM , it is interesting to see how a TeV-scale L-R seesaw can be realized in Nature. There are essentially three ways to do so: (i) by choosing one set of Yukawa couplings to be $\lesssim 10^{-11/2}$ for a particular VEV assignment for the bidoublet Higgs fields; (ii) by choosing particular symmetry-motivated textures for the Yukawa couplings which ensure that the leading order seesaw contribution to neutrino masses vanish; and (iii) by choosing large Yukawa couplings as allowed by specific textures and invoking cancellations in the seesaw formula \eqref{Mnu}. The strategy (i) with small Yukawa couplings does not lead to any new interesting phenomenology over the minimal LRSM. Therefore, we will focus on strategies (ii) and (iii) in the following sections, and in particular, use strategy (iii) for our numerical analysis for the calculation of lepton asymmetry. A similar analysis can be performed following strategy (ii) as well. For the phenomenological implications of this new class of L-R seesaw models, see e.g.~\cite{DLM, CDM}. 

\subsection{A symmetry-protected L-R seesaw with large light-heavy neutrino mixing}\label{sec2.1}
As mentioned above, the basic strategy for understanding small neutrino masses with a low-scale  type-I seesaw is to have the appropriate textures for $M_D$ and $M_N$ in Eq.~\eqref{Mnu}. 
There are several examples of this type~\cite{kersten, texture} in the context of the minimal seesaw extensions of the SM. Here, we discuss the embedding of one such texture (from Ref.~\cite{kersten}) in the L-R model using an appropriate family symmetry~\cite{DLM}. The symmetry must not only guarantee the special leptonic textures but also must be free of light scalar bosons which can result if the effect of the discrete symmetry is to automatically lead to a $U(1)$ symmetry. Moreover in LRSM, the charged-lepton mass matrix $M_l$ is related to $M_D$ which puts additional constraints on phenomenological viability of the model. Therefore, we find it remarkable that the model presented below remains a viable TeV-scale L-R type-I seesaw model for neutrinos, and as a result, has interesting phenomenological implications beyond the minimal LRSM.

The Dirac and Majorana mass matrices $M_D$ and $M_N$ considered here have the following form:
\begin{eqnarray}
M_D \ = \ \left(\begin{array}{ccc} m_1 & \delta_1 & \epsilon_1\\ m_2 & \delta_2 & \epsilon_2\\ m_3 &\delta_3 & \epsilon_3\end{array}\right), \qquad \qquad 
M_N\ = \ \left(\begin{array}{ccc} \delta M & M_1 & 0\\M_1 & 0 & 0 \\ 0&0&M_2\end{array}\right),
\label{eq:texture}
\end{eqnarray}
with $\epsilon_i, \delta_i \ll m_i $ and $\delta M  \ll M_i$. In the limit of $\epsilon_i, \delta_i, \delta M \to 0$, the neutrino masses vanish, although the heavy-light mixing parameters given by $\xi_{ij}=m_i/M_j$ can be quite large. The light neutrino masses given by the seesaw formula \eqref{Mnu} become proportional to either $\epsilon_i$, $\delta_i$ or their products. If the smallness of $\delta_i$ and $\epsilon_i$ can be guaranteed by some symmetry, then  we have a ``natural'' TeV-scale seesaw model with tiny neutrino masses.  As we show below, the mass textures~\eqref{eq:texture} can be successfully embedded into the L-R framework, while satisfying all current experimental constraints. Note that the choice of texture for the RH neutrino mass matrix in Eq.~\eqref{eq:texture} is necessary to keep the model compatible with strong bounds from tree-level $\mu^-\to e^-e^+e^-$ decay via TeV-scale $\Delta_R$ exchange~\cite{DLM2}.

In order to obtain the special Dirac and Majorana textures given in Eq.~\eqref{eq:texture}, we supplement the L-R gauge group $SU(2)_L\times SU(2)_R\times U(1)_{B-L}$ with a global discrete symmetry $D\equiv Z_4\times Z_4\times Z_4$~\cite{DLM}. For the Higgs sector, we choose three bi-doublets ($\phi_{1,2,3}$) with $B-L=0$ and two RH triplets ($\Delta_{R,1}, \Delta_{R,2}$) with $B-L=2$. The fermion and Higgs multiplets are assigned the $D$ quantum numbers as shown in Table~\ref{tab1}. The leptonic part of the Yukawa Lagrangian \eqref{yuk} invariant under the $D$-symmetry is given by
\begin{eqnarray}
-{\cal L}_{l,Y} \ = \ h_{i1}\bar{L}_i \tilde{\phi}_1R_1+h_{i2} \bar{L}_i\phi_2R_2+h_{i3}\bar{L}_i\phi_3R_3 
+ f_{12}R_1R_2\Delta_{R,1}+f_{33}R_3R_3\Delta_{R,2}+{\rm H.c.}
\end{eqnarray}
In the symmetry limit, the VEVs of $\phi_{2,3}$ have the form $\langle \phi_{a}\rangle = {\rm diag}(0,\kappa'_{a})$. This is because the 
terms of the form Tr$ (\tilde{\phi}_{a}\phi^\dagger_{b})$ which would change the $\phi$ VEV to the form diag$ (\kappa, \kappa')$ are forbidden by the $D$-symmetry from appearing in the scalar potential.
Thus in the symmetry limit, the $M_D$ elements in one column are big and non-zero, and the charged lepton mass matrix has one eigenvalue zero which can be identified as the electron flavor. The RH neutrino Majorana mass matrix has the form in Eq.~(\ref{eq:texture}) with $\delta M=0$. To make the model realistic, we add small soft symmetry-breaking terms of the form 
\begin{align}
\delta V(\phi) \ = \  \sum_{a,b}\mu^2_{ab}{\rm Tr}(\tilde{\phi}_{a}\phi^\dagger_b) + {\rm H.c.}
\end{align}
 to the scalar potential. 
This will induce the $\phi_{a}$ VEVs of the form $\langle \phi_{a}\rangle = {\rm diag}(\delta \kappa_a,\kappa'_{a})$, where 
\begin{align}
\delta \kappa_a \ \propto \ \frac{\sum_b\mu^2_{ab}\kappa'_b}{\sum_{a,b}\lambda'_{ab}\kappa^\prime_a\kappa^\prime_b+\sum_a\lambda_a v^2_{R,a}} \; ,
\end{align}
 with $\lambda,\lambda'$ respectively being the scalar self-couplings of the bidoublet and triplet fields in the scalar potential. Choosing appropriately small $\mu^2_{ab}$, we can get very small $\delta \kappa_{a}$ as required to satisfy the neutrino oscillation data. 
\begin{table}[t]
\begin{center}
\begin{tabular}{c|c}\hline\hline
Field & $Z_4\times Z_4\times Z_4$ transformation\\ \hline
$L_{1,2,3}$ & $(1,~1,~1)$\\
$R_1$ & $(-i,~1,~1)$ \\
$R_2$ & $(1,~-i, ~1)$ \\
$R_3$ & $(1,~1,~-i)$ \\
$\phi_1$ & $(-i,~1,~1)$\\
$\phi_2$ & $(1,~i,~1)$\\
$\phi_3$ & $(1,~1,~i)$\\
$\Delta_{R,1}$ & $(i, ~i,~1)$\\
$\Delta_{R,2}$ & $(1, ~1,~-1)$\\
\hline\hline
\end{tabular}
\end{center}
\caption{The discrete symmetry assignments for the fermion and Higgs fields in our L-R model that lead naturally to the special Dirac and Majorana textures given in Eq.~(\ref{eq:texture})~\cite{DLM}.}\label{tab1}
\end{table}

\subsection{Naturalness of small $\delta \kappa$} \label{natkappa}
It is possible to generate the small parameters $\delta \kappa_a$ naturally through loop effects involving the $W_L-W_R$ mixing. 
To explain this, first we note that if in a L-R model, $\delta \kappa$'s are set to zero in a natural manner, it will lead to zero $W_L-W_R$ mixing. By the same token, when $W_L-W_R$ mixing is nonzero, it will induce a nonzero $\delta \kappa$. Thus there is an intimate connection between $\delta \kappa$ and $W_L-W_R$ mixing. In our model discussed above, $\delta \kappa=0$ is guaranteed by $(Z_4)^3$ discrete symmetry. However,  if bi-doublets are used to generate quark masses, it is not possible to set either $\kappa$ or $\kappa'$ to zero, since in that case the symmetry that stabilizes the vanishing of $\kappa$ or $\kappa'$ makes it impossible to get realistic quark masses and mixing. Once in the quark sector both $\kappa$ and $\kappa'$ are set to nonzero values, they will induce tree level $W_L-W_R$ mixing and hence arbitrary $\delta \kappa\neq 0$ at the tree level in the leptonic sector, thus invalidating the naturalness of the whole scenario. There is however an alternative way to get quark flavor pattern right without introducing bidoublet Higgs fields and thereby avoiding nonzero $W_L-W_R$ mixing at the tree level.
If we introduce $SU(2)_{L,R}$ singlet vector-like up (${\cal U}$) and down (${\cal D}$) quarks, and only LH and RH doublet Higgs fields $\chi_{L,R}$ with nonzero VEVs which give mass to the quarks via a quark seesaw mechanism as in~\cite{quarkseesaw}, it keeps the $\delta \kappa$'s naturally small. To briefly introduce the quark seesaw, we write the Yukawa interactions responsible for fermion masses in this model as 
\begin{eqnarray}
- \mathcal{L}_Y \ = \ 
 Y_u  \bar{Q}_L \tilde{\chi}_L {\cal U}_R
+ Y_d \bar{Q}_L  \chi_L {\cal D}_R
 + (L \leftrightarrow R) 
+ \bar{\cal U}_L M_{\cal U} {\cal U}_R
+ \bar{\cal D}_L M_{\cal D} {\cal D}_R
 + {\rm H.c.} \,.
\end{eqnarray}
where $\tilde{\chi}_{L,R} = i\tau_2\chi^*_{L,R}$.
This leads to 
generic seesaw type quark mass relations, e.g. 
\begin{eqnarray}
\label{seesaw}
M_{u} \ \simeq \ -\frac{Y^2_u v_L v_R}{2M_{\cal U}} \,.
\end{eqnarray}
 The leptonic sector bidoublets do not couple to the quark sector due to the $(Z_4)^3$ discrete symmetry under which the quark sector fields are singlets. As a result, the $W_L-W_R$ mixing is  zero also in the quark sector,  and hence, $\delta \kappa$ is not induced at the tree level. However, $W_L-W_R$ mixing is induced at the one loop level via the exchange of top and bottom quarks in the quark sector (see Figure 1), which in turn induces leptonic $\delta \kappa$ at the two loop level from the gauge interaction $g_Lg_R\phi^{0\dagger}_1\phi^0_2W^+_LW^-_R$ in the notation of Eq.~\eqref{LRhiggs}. This makes $\delta \kappa$ finite and small, thus keeping the model technically natural.
 \begin{figure}[t!]
\centering
\includegraphics[width=7cm]{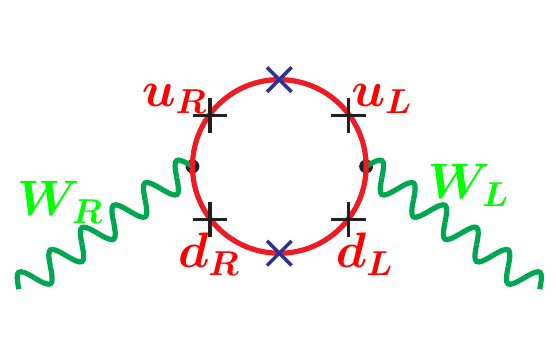}
\caption{One loop graph for $W_L-W_R$ mixing in the quark seesaw model. The (blue) crosses in the middle of the two quark propagators are for the vectorlike quark mass insertions, while the other four (black) crosses denote the mixing between the vectorlike quarks and SM quarks.}
\label{kappa}
\end{figure}

An important point to notice is that $\delta \kappa_a$'s are responsible for the electron mass as well as neutrino masses via type-I seesaw. Thus getting a fit to the observed neutrino masses and mixing while at the same time keeping electron mass at its desired value is a nontrivial task since in the lepton sector, the model has only 14 free parameters (9 Yukawa couplings, 3 Majorana mass parameters and two VEVs) to start with. One can always go to a basis where $M_D$ takes an upper-triangular form, in which case the number of free parameters reduces to 11. Out of this 11 model parameters, we must not only get fits for the three charged lepton masses, the two neutrino mass-squared differences and three mixing angles (total of 8 outputs), but must also satisfy the unitarity constraints on the new light neutrino mixing matrix as well as constraints from rare lepton decays which involve only the RH mixing matrix for charged leptons. 
It was explicitly shown in~\cite{DLM} that this model does indeed provide a fit to all observables while satisfying all the necessary constraints.

\section{An alternative neutrino mass fit}\label{sec3}
In this section, we discuss neutrino fit in a model with $M_D$ and $M_N$ texture as given in Eq.~\eqref{eq:texture} without worrying about its origin from discrete symmetries and use this fit in the subsequent section for the discussion of leptogenesis. We consider a minimal scenario with only one bidoublet $\phi$ and a single right-Higgs triplet $\Delta_R$ to illustrate the fit. Clearly, in this case we will have to adjust parameters to get the desired fermion mass textures. 

In this class of L-R models, we can always choose a basis of the LH sector prior to $SU(2)_L\times U(1)_Y$ breaking such that the Dirac mass matrix $M_D$ can be written in an upper-triangular form without affecting the RH neutrino texture. In this basis, the mass matrices
obtained from the Yukawa Lagrangian in Eq.~(\ref{yuk}) are given by
\begin{align}
	M_l \ =\ & \left(\begin{array}{ccc}
		h_{11} \kappa' + \tilde{h}_{11} \kappa^* & h_{12} \kappa' + \tilde{h}_{12} \kappa^* & h_{13} \kappa' + \tilde{h}_{13} \kappa^* \\
		\frac{|\kappa'|^2 - |\kappa|^2}{\kappa'^*} h_{21} & h_{22} \kappa' + \tilde{h}_{22} \kappa^* & h_{23} \kappa' + \tilde{h}_{23} \kappa^* \\
		\frac{|\kappa'|^2 - |\kappa|^2}{\kappa'^*} h_{31} & \frac{|\kappa'|^2 - |\kappa|^2}{\kappa'^*} h_{32} & h_{33} \kappa' + \tilde{h}_{33} \kappa^*
	\end{array} \right), \\
	M_D \ = \ & \left(\begin{array}{ccc}
		h_{11} \kappa + \tilde{h}_{11} \kappa'^* & h_{12} \kappa + \tilde{h}_{12} \kappa'^* & h_{13} \kappa + \tilde{h}_{13} \kappa'^* \\
		0 & h_{22} \kappa + \tilde{h}_{22} \kappa'^* & h_{23} \kappa + \tilde{h}_{23} \kappa'^* \\
		0 & 0 & h_{33} \kappa + \tilde{h}_{33} \kappa'^*
	\end{array} \right), \label{eq:MD2} \\
	M_N  \ = \ & \left(\begin{array}{ccc}
		\delta M & f_{12} v_{R,1} & 0 \\
		f_{12} v_{R,1} & 0 & 0 \\
		0 & 0 & 2 f_{33} v_{R,2}
	\end{array} \right), \label{eq:MN2}
\end{align}
Note that since parity is a broken symmetry in this low energy version, we need not have the Yukawa couplings $h_{ij}$ to satisfy hermiticity.
We choose the parameters of the model, i.e. $h_{ij}$'s and $\kappa$, $\kappa'$ such that all the experimental constraints in the lepton sector are satisfied. As an example fit, we consider the mass matrices of the following form:
\begin{align}
	M_l \ = \ & \left(\begin{array}{ccc}
		0.00120 & -0.0507 & -1.41 \\
		0 & 0.0929 & -0.657 \\
		0 & 0 & -0.861
	\end{array} \right) \text{GeV},
	\label{eq:Menum} \\
	M_D \ = \ & \left(\begin{array}{ccc}
		0.0676 \ e^{i 0.557 \pi} & -1.06 \times 10^{-7} \ e^{-i 0.557 \pi} & -1.06 \times 10^{-4} \\
		0 & 1.95 \times 10^{-7} \ e^{-i 0.557 \pi} & -4.78 \times 10^{-5} \\
		0 & 0 & -6.38 \times 10^{-5}
	\end{array} \right) \text{GeV},
	\label{eq:MDnum} \\
	M_N \ = \ & \left(\begin{array}{ccc}
		2.59 \times 10^{-6} & 585 & 0 \\
		585 & 0 & 0 \\
		0 & 0 & -585
	\end{array} \right) \text{GeV},
	\label{eq:MNnum}
\end{align}
In addition, we choose the VEVs $\kappa = 6.46$ GeV and $\kappa' = 173.88~e^{i 0.700 \pi}$ GeV, which satisfy $|\kappa^2| + |\kappa'^2| = v^2$ with $v$ = 174 GeV. These are all the input parameters for the model. The $M_l$ is diagonalized by a bi-unitary transformation: $\widehat{M}_l = (V_l^L)^\dagger M_l V_l^R$, where 
\begin{align}
	V_l^L = i \left(\begin{array}{ccc}
		0.426 & 0.436 & 0.793 \\
		0.232 & -0.900 & 0.369 \\
		-0.875 & -0.0271 & 0.484
	\end{array} \right), \quad
	V_l^R = i \left(\begin{array}{ccc}
		1.00 & 0.00495 & 0.000535 \\
		0.00495 & -1.00 & -0.00332 \\
		0.000519 & 0.00332 & -1.00
	\end{array} \right).
	\label{Vl}
\end{align}
The light neutrino mass matrix $M_{\nu}$ is diagonalized by a unitary transformation: $\widehat{M}_\nu = V_\nu^{\sf T} M_\nu V_\nu$, where 
\begin{align}
	V_\nu \ = \ \left(\begin{array}{ccc}
		0.338 i & -0.0149 & 0.941 \\
		0.742 i & 0.619 & -0.257 \\
		-0.579 i & 0.785 & 0.221
	\end{array} \right).
\end{align}
The Pontecorvo-Maki-Nakagawa-Sakata (PMNS) mixing matrix is then given by 
\begin{align}
	V_\text{PMNS} \ = \ (V_l^{L})^\dagger V_\nu 
	\ = \  \left(\begin{array}{ccc}
		0.823 & 0.549 & 0.148 \\
		-0.505 & 0.585 & 0.635 \\
		0.262 & -0.597 & 0.758
	\end{array} \right)
	\left(\begin{array}{ccc}
		1 & 0 & 0 \\
		0 & i & 0 \\
		0 & 0 & -i
	\end{array} \right),
	\label{pmns}
\end{align}  
The resulting neutrino masses and mixing angles as well as the charged-lepton masses are given in Table \ref{tab:fit}.
\begin{table}[t]
	\begin{center}
		\begin{tabular}{c|c} \hline \hline
			Parameter & Value \\ \hline
			$m_e$ & $0.511$ MeV  \\
			$m_\mu$ & $106$ MeV \\
			$m_\tau$ & $1.78$ GeV \\ \hline
			$m_{\nu_1}$ & $6.49 \times 10^{-3}$ eV \\
			$m_{\nu_2}$ & $1.08 \times 10^{-2}$ eV \\
			$m_{\nu_3}$ & $5.02 \times 10^{-2}$ eV \\
			\hline
			$\theta_{12}$ & $33.7^\circ$ \\
			$\theta_{23}$ & $40.0^\circ$ \\
			$\theta_{13}$ & $8.51^\circ$ \\ \hline
                        	$m_{N_1}$ & $585$ GeV \\
			$m_{N_2}$ & $585$ GeV \\
			$m_{N_3}$ & $585$ GeV \\ \hline
			\hline
		\end{tabular}
	\end{center}
	\caption{Masses and mixing angles calculated from the mass matrices given by Eqs.~(\ref{eq:Menum})-(\ref{eq:MNnum}). The mass splitting between the RH neutrinos is of order of $\delta M\sim 10^{-6}$ GeV. }
	\label{tab:fit}
\end{table}

Since $M_l = h \kappa' + \tilde{h} \kappa^*$ and $M_D = h \kappa + \tilde{h} \kappa'^*$, the Yukawa couplings $h,~\tilde{h}$ are easily calculated using VEV's chosen as above, and they are found to be
\begin{align}
	h & \ = \ \left(\begin{array}{ccc}
		-(0.15 + 1.98 i) \times 10^{-5} & (1.72 + 2.36 i) \times 10^{-4} & (4.77 + 6.57 i) \times 10^{-3} \\
		0 & -(3.14 + 4.33 i) \times 10^{-4} & (2.22 + 3.06 i) \times 10^{-3} \\
		0 & 0 & (2.91 + 4.01 i) \times 10^{-3}
	\end{array} \right), \label{eq:yuk} \\
	\tilde{h} & \ = \  \left(\begin{array}{ccc}
		-(2.69 + 2.81 i) \times 10^{-4} & 1.08 \times 10^{-5} & 3.02 \times 10^{-4} \\
		0 & -1.99 \times 10^{-5} & 1.41 \times 10^{-4} \\
		0 & 0 & 1.84 \times 10^{-4}
	\end{array} \right). \label{eq:yukt}
\end{align}
The leptonic $\CP$ asymmetry generated by the RH neutrino decays is governed by these Yukawa couplings. If we assume that the two SM doublets in the LR bi-doublet i.e. $\phi_1$ and $\phi_2$ are approximate mass eigenstates with $m_{\phi_1} > m_N>m_{\phi_2}$, the dominant contribution to the $\CP$ asymmetry  comes solely from the self-energy correction by the $\phi_2$-loop to the decay process $N \to \phi_2 L_l$. Therefore, only the Yukawa coupling matrix $\tilde{h}$ determines the flavor effect in RH neutrino interactions relevant to resonant leptogenesis. 

The RH neutrino mixing effect at resonance can be captured by the one-loop resummed Yukawa couplings~\cite{Pilaftsis:2003gt}. Explicitly, in the heavy-neutrino mass eigenbasis, we obtain for the resummed Yukawa couplings $\widehat{\maf{h}}_{l\alpha}$ and their $\CP$-conjugates $\widehat{\maf{h}}^c_{l\alpha}$ the following values: 
\begin{align}
	\widehat{\maf{h}} & \ = \ \left(\begin{array}{ccc}
		(-2.52 + 1.03 i) \times 10^{-4} & (2.39 + 1.00 i) \times 10^{-4} & (-0.03 + 3.46 i) \times 10^{-4} \\
		-(2.82 + 5.07 i) \times 10^{-5} & (4.37 - 3.68 i) \times 10^{-5} & (-0.08 + 1.49 i) \times 10^{-4} \\
		-(3.63 + 4.69 i) \times 10^{-5} & (3.77 - 4.89 i) \times 10^{-5} & (0.00 + 1.84 i) \times 10^{-4}
	\end{array} \right), \label{eq:resyuk} \\
	\widehat{\maf{h}}^c & \ = \ \left(\begin{array}{ccc}
		-(1.59 + 1.26 i) \times 10^{-4} & (1.40 - 1.25 i) \times 10^{-4} & (1.02 - 7.03 i) \times 10^{-5} \\
		(2.71 + 4.90 i) \times 10^{-5} & (-1.57 + 3.67 i) \times 10^{-5} & (0.02 - 1.39 i) \times 10^{-4} \\
		(3.62 + 4.70 i) \times 10^{-5} & (-3.78 + 4.88 i) \times 10^{-5} & (0.00 + 1.84 i) \times 10^{-4}
	\end{array} \right). \label{eq:resyukc}
\end{align}
The heavy neutrino decay rates for the Yukawa-mediated two-body processes $N_\alpha\to L_l\phi,~L_l^c\phi^c$ are calculated with these resummed Yukawa couplings: 
\begin{align}
  \Gamma(N_{\alpha}\to L_l\phi) \
  = \  \frac{m_{N_\alpha}}{16\pi}{\widehat{\maf{h}}}_{l \alpha}{\widehat{\maf{h}}}^*_{l \alpha}\; , \qquad   \Gamma(N_{\alpha}\to L^c_l\phi^c ) \
  = \  \frac{m_{N_\alpha}}{16\pi}{\widehat{\maf{h}}^c}_{l \alpha}{\widehat{\maf{h}}}^{c*}_{l \alpha}\; .
  \label{gamma}
\end{align}
which are crucial parameters for the calculation of lepton asymmetry, as we will see in the next section. 

\section{Resonant leptogenesis with TeV-scale $W_R$}\label{sec4} 
As eluded to above, for low scale seesaw models, a simple way to generate enough lepton asymmetry is to use resonant leptogenesis~\cite{Pilaftsis:1997jf, Pilaftsis:2003gt}. It requires at least two RH neutrinos to be nearly degenerate, which is guaranteed in our model by the choice of the texture in the RH neutrino mass matrix given by Eq.~\eqref{eq:MN2}. Thus, in addition to guaranteeing suppression of lepton flavor violating processes such as $\mu\to 3e$, this choice of RH neutrino texture also provides a necessary condition for resonant leptogenesis. As far as the Dirac neutrino mass matrix $M_D$ is concerned, we choose the pattern given in Eq.~\eqref{eq:MD2} so that it fits neutrino oscillation data. Since our focus is on the feasibility of low scale leptogenesis in a realistic model for neutrinos, we do not stress here on the naturalness of the texture, although a similar exercise can be performed with the fit presented in Section~\ref{sec2.1}. The details of leptogenesis in this class of models was discussed in~\cite{DLM2} and we do not repeat the basic equations here. However, we have extended the work of~\cite{DLM2} and also present here some revised numerical results due to a computational error in the code used there.

The basic steps in our analysis are as follows: first we calculate the flavored $\CP$ asymmetry
\begin{align}
\varepsilon_{l \alpha} \ = \ \frac{1}{\Gamma_{N_\alpha}} [\Gamma (N_\alpha \to L_l \phi) - \Gamma (N_\alpha \to L_l^c \phi^c)] \; ,
\label{epsilon}
\end{align}
where the partial decay widths are given by Eq.~\eqref{gamma} and the total decay width is given by 
\begin{align}
\Gamma_{N_\alpha} \ = \ \sum_l\left[\Gamma(N_\alpha \to L_l \phi) + \Gamma(N_\alpha\to L_l^c\phi^c) \right] + 2\: \Gamma(N_\alpha\to l_R q_R\bar{q}'_R) \; .
\label{tot_width}
\end{align}
Note that in LRSM, there is an additional contribution to the total decay width due to the three-body decay mediated by $W_R$: 
\begin{eqnarray}
\Gamma(N_\alpha\to l_R q_R\bar{q}'_R) \ = \ \Gamma(N_\alpha\to \bar{l}_R \bar{q}_R q'_R) \ = \ \frac{3g_R^4}{2^9\pi^3 m_{N_\alpha}^3} \int_0^{m_{N_\alpha}^2} ds \frac{m_{N_\alpha}^6-3m_{N_\alpha}^2 s^2+2 s^3}{(s-M_{W_R}^2)^2+M_{W_R}^2\Gamma_{W_R}^2} \; ,
\label{3body}
\end{eqnarray}
where $\Gamma_{W_R}\simeq (g_R^2/4\pi)M_{W_R}$ is the total decay width of $W_R$, assuming that all three heavy neutrinos are lighter than $W_R$. The $\CP$ asymmetry~\eqref{epsilon} in our case is determined by the RH neutrino texture, specifically, the magnitudes of $M_{1,2}$ and $\delta M$ in Eq.~\eqref{eq:texture} as well as the magnitudes of the Yukawa couplings, and these parameters are constrained by the neutrino fit. 

The second step in our calculation is to calculate the thermodynamic evolution of the normalized heavy neutrino and lepton doublet number densities $\eta^N_\alpha$ and $\eta_l^{\Delta L}$ respectively, where $\eta^X\equiv n^X/n^\gamma$ and $n^\gamma=2m_{N_1}^3\zeta(3)/(\pi^2 z^3)$ is the photon number density, $\zeta(x)$ being the Riemann zeta function. For simplicity and for a fair comparison with the results of~\cite{hambye}, here we use the flavor-diagonal Boltzmann equations~\cite{Deppisch:2010fr}\footnote{Including flavor off-diagonal effects could lead to additional enhancement of the final lepton asymmetry~\cite{pilaf2, Dev:2014wsa}.}:
\begin{align}
\frac{d\eta^N_\alpha}{dz} \ & = \ -\left(\frac{\eta^N_\alpha}{\eta^N_{\rm eq}}-1\right)(D_\alpha+S_\alpha) \;, \label{be1} \\
\frac{d\eta^{\Delta L}_l}{dz} \ & = \ \sum_\alpha \varepsilon_{l\alpha}\left(\frac{\eta^N_\alpha}{\eta^N_{\rm eq}}-1\right)\widetilde{D}_\alpha -\frac{2}{3} \eta^{\Delta L}_l W_l \;, \label{be2}
\end{align}
where 
$z=m_{N_1}/T$ is a dimensionless variable ($T$ being the temperature of the Universe) and
$\eta^N_{\rm eq}  \equiv  n^N_{\rm eq}/n^\gamma  =  z^2 K_2(z)/2\zeta(3) $
is the heavy neutrino equilibrium number density, $K_n(x)$ being the $n$-th order modified Bessel function of the second kind.  The various decay ($D_\alpha,~\widetilde{D}_\alpha$), scattering ($S_\alpha$) and washout ($W_l$) rates appearing in Eqs.~\eqref{be1} and \eqref{be2} are given by 
\begin{align}
		\widetilde{D}_\alpha & \ = \ \frac{z}{n^\gamma H_N} \sum_k \widetilde{\gamma}^D_{k \alpha}, \label{Dtilde} \\
	D_\alpha & \ = \ \frac{z}{n^\gamma H_N} \sum_k \gamma^D_{k \alpha}, \\
	S_\alpha & \ = \ \frac{z}{n^\gamma H_N} \sum_k (\gamma^{S_L}_{k \alpha} + \gamma^{S_R}_{k \alpha}) ,\\
W_l  & \ = \ \frac{z}{n^\gamma H_N} \left[ \sum_\alpha \left( B_{l \alpha} \sum_k \gamma^D_{k \alpha} + \widetilde{\gamma}^{S_L}_{l \alpha} + \widetilde{\gamma}^{S_R}_{l \alpha} \right) + \sum_k \left( \gamma^{(\Delta L = 2)}_{lk} + \gamma^{(\Delta L = 0)}_{lk} \right) \right] \nonumber \\
	& \ \equiv \  \frac{1}{2 \zeta (3)} z^3 K_1 (z) K_l^\text{eff} (z) \; , \label{wash}
\end{align}
where $H_N\equiv H(z=1)\simeq 17 m_{N_1}^2/M_{\rm Pl}$ is the Hubble parameter at $z=1$, assuming only SM degrees of freedom in the thermal bath, $M_{\rm Pl}=1.2\times 10^{19}$ GeV is the Planck mass and $B_{l\alpha}$ is the branching fraction of the RH neutrino decays relevant for the generation of $\CP$ asymmetry: 
\begin{align}
		B_{l \alpha} \ = \ \frac{1}{\Gamma_{N_\alpha}} [\Gamma (N_\alpha \to L_l \phi) + \Gamma (N_\alpha \to L_l^c \phi^c)] \; . 
\label{BR}
\end{align}
The various $\gamma$'s appearing in Eqs.~\eqref{Dtilde}-\eqref{wash} represent the reaction rates that involve the RH neutrino decays and inverse decays as well as other $2\leftrightarrow 2$ scattering processes in the model (see~\ref{app}). Only the two-body decays of the RH neutrinos involving complex Yukawa couplings are responsible for building up the asymmetry, whereas all other processes lead to washout effects. 

\begin{figure}[t]
	\centering
		\includegraphics[width = 7.5 cm]{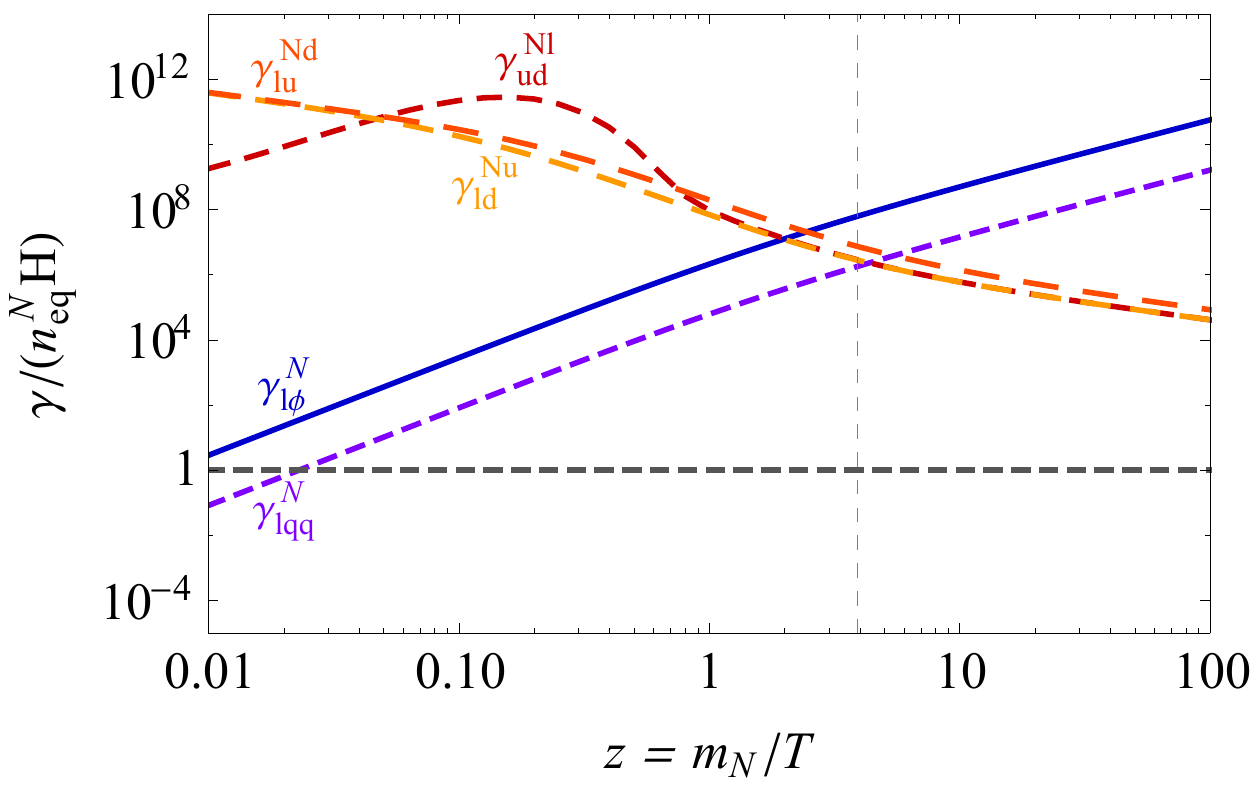} 
		\includegraphics[width = 7.5 cm]{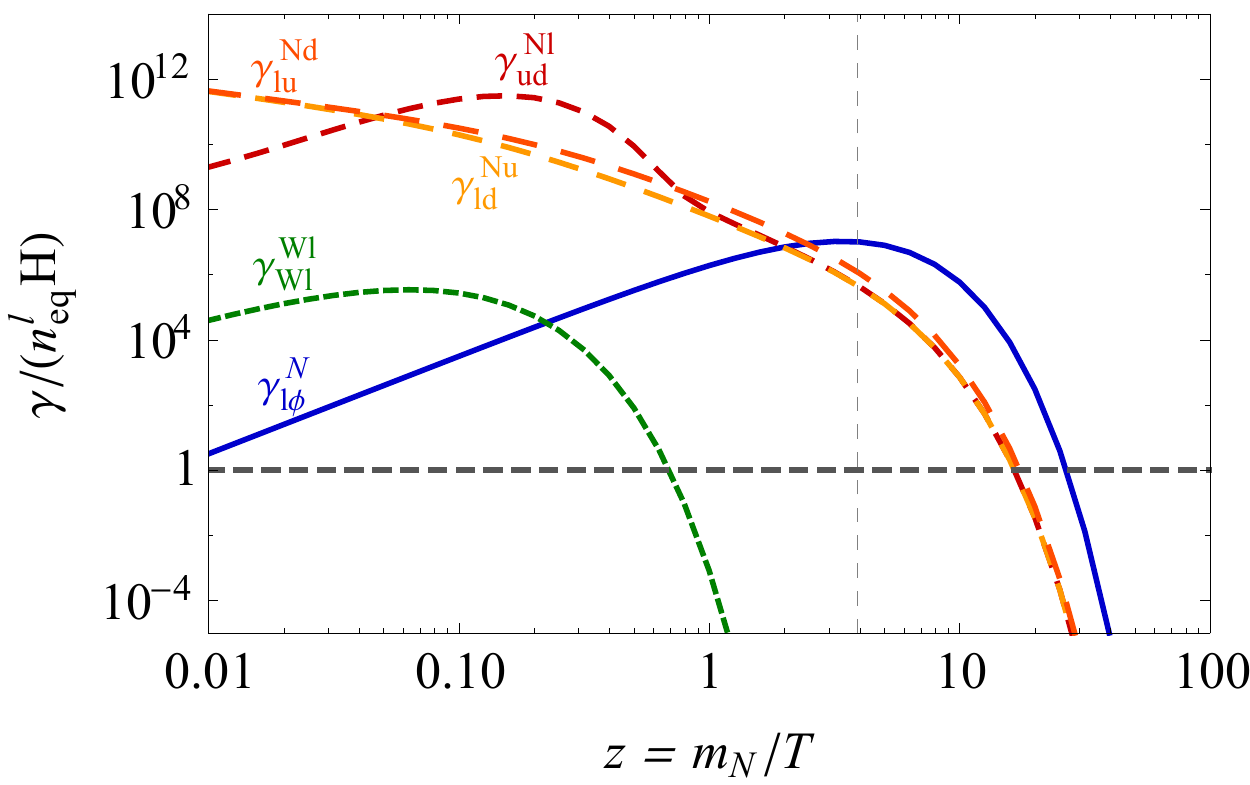} 
		\caption{Evolution of $\gamma / (n^N_\text{eq} H)$ [left panel] and $\gamma / (n^l_\text{eq} H)$ [right panel]. For the latter, we have also included the scattering rate of $W_R^+ l_R^- \to W_R^- l_R^+$ with $m_{\Delta^{++}}$ = 10 TeV and $m_{W_R}=13.1$ TeV. }
	\label{fig:gnH}
\end{figure}

We present in Figure~\ref{fig:gnH} the evolution of the various collision rates as given in~\ref{app}. The left panel of the Figure is relevant to the first Boltzmann equation \eqref{be1}, while the second one is relevant to \eqref{be2}. The vertical dashed line shows the critical value $z=z_c$ beyond which the sphaleron processes become ineffective. The horizontal dashed line is shown for easy comparison with the Hubble rate. Here we have used $m_{W_R}=13.1$ TeV which is the lowest value of $m_{W_R}$ we found in this model with an {\em explicit} neutrino fit that satisfies the leptogenesis constraints. 

\begin{figure}[t]
	\centering
		\includegraphics[width = 7.5 cm]{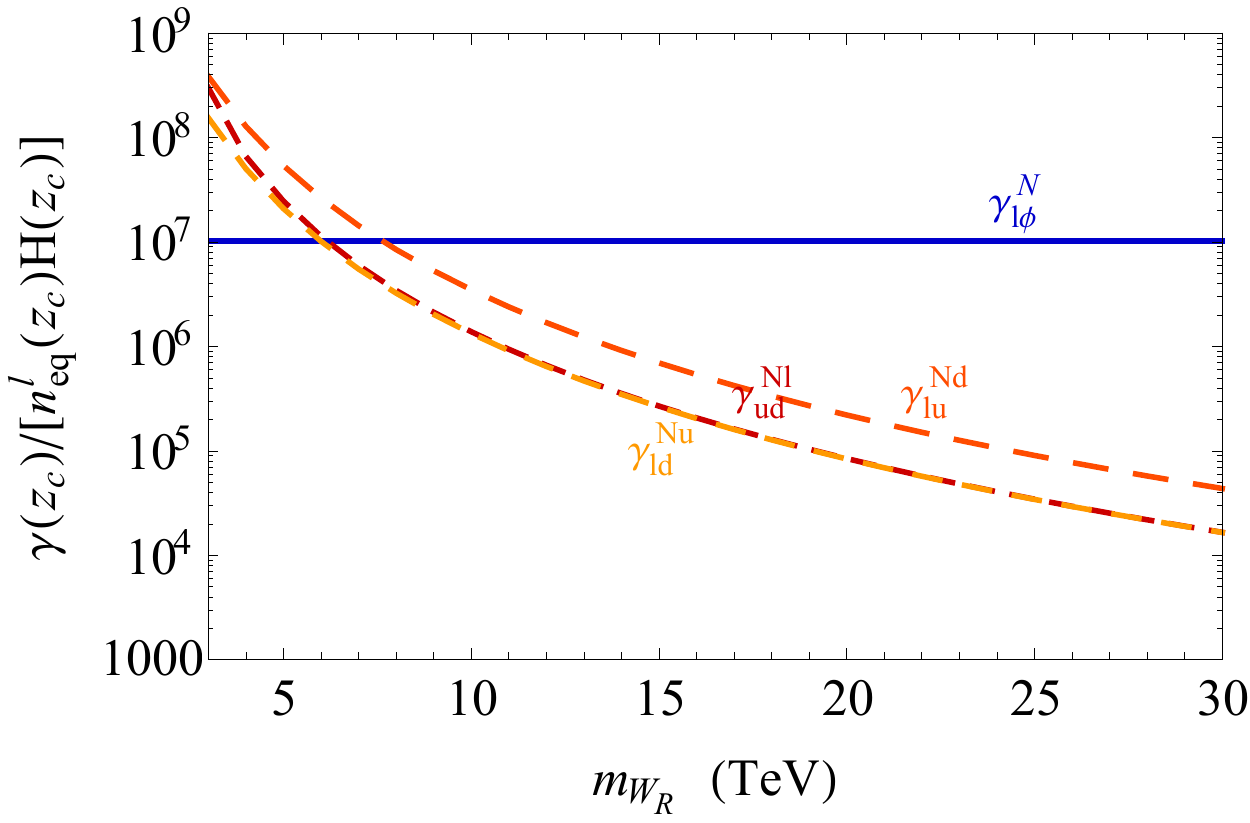} 
		\includegraphics[width = 7.5 cm]{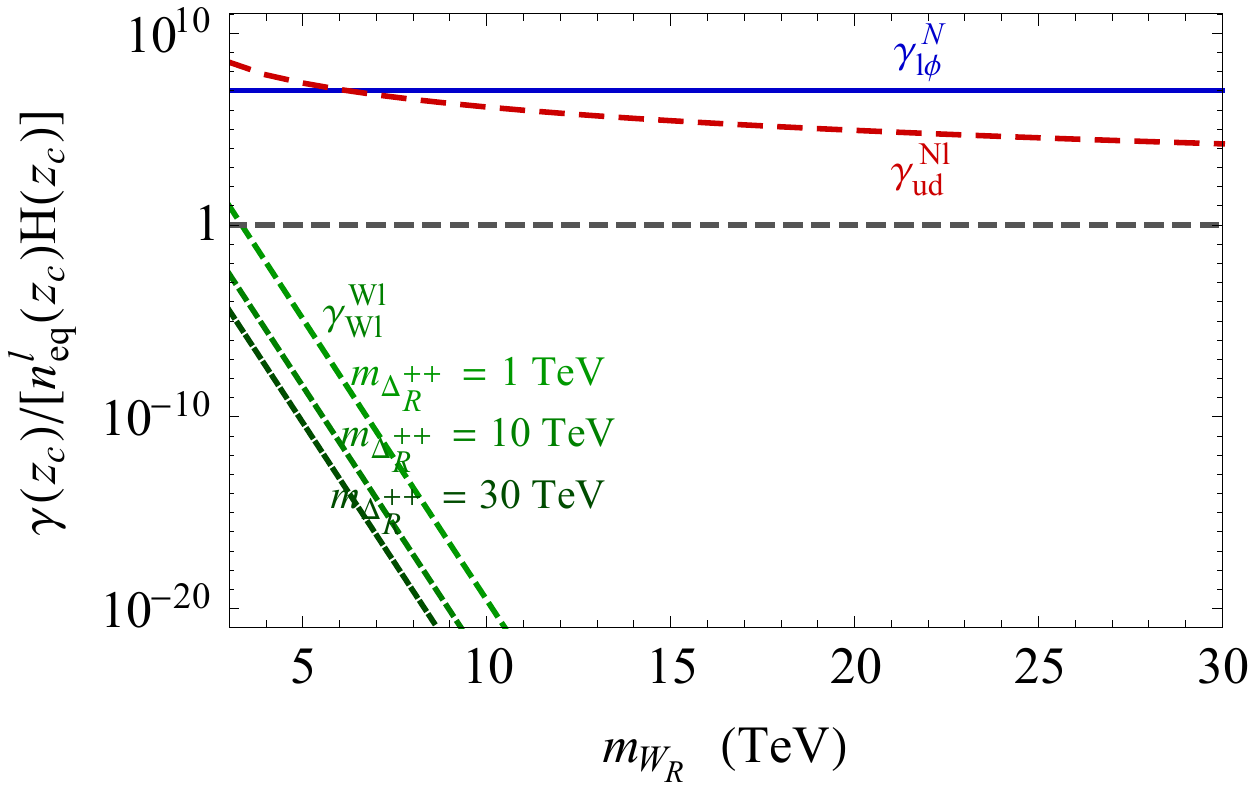}
		\caption{Various reaction rates $\gamma (z_c) / [n_l^\text{eq} (z_c) H (z_c)]$ as a function of $m_{W_R}$ for $m_N$ = 585 GeV. The green lines in the right panel correspond to the process $W^+_Rl^-_R\to W^-_Rl^+_R$ for various $\Delta^{++}_R$ masses.}
	\label{fig:glHWe}
\end{figure}

Recently, it was pointed out~\cite{sarkar} that the scattering process $W_R^+ l_R^- \to W_R^- l_R^+$ mediated by the doubly-charged component of the RH triplet $\Delta_R^{++}$ might spoil leptogenesis, if $m_{W_R}$ and $m_{\Delta_R^{++}}$ are relatively small. We have explicitly calculated the washout effect due to this process for various values of $m_{W_R}$ and $m_{\Delta^{++}}$. As shown in Figure~\ref{fig:glHWe} and also in Figure~\ref{fig:gnH} (right panel), this effect is negligible compared with the decay rate and other scattering rates when $m_{W_R}$ and $m_{\Delta^{++}}$ are in the regime of interest. 

In the strong washout regime, the flavored lepton asymmetry is approximately given by~\cite{Buchmuller:2004nz, DLM2}
\begin{align}
	\eta^{\Delta L} (z)_{l\alpha} \ \simeq \  \frac{3}{2 z K_l^\text{eff} (z)} \varepsilon_{l \alpha} \frac{\widetilde{D}_\alpha (z)}{D_\alpha (z) + S_\alpha (z)} \; ,
	\label{eq:etaL}
\end{align}
where $K_l^{\rm eff}$ is the effective washout parameter as defined in Eq.~\eqref{wash} and $\widetilde{D}/(D+S)$ is the effective dilution factor.  
For the numerical fit given in Section~\ref{sec3}, we obtain
 \begin{align}
	\eta^{\Delta L}_{l \alpha} = \left(\begin{array}{ccc}
		5.16 \times 10^{-9} & 5.13 \times 10^{-9} & 1.32 \times 10^{-8} \\
		6.44 \times 10^{-11} & 4.89 \times 10^{-10} & 6.48 \times 10^{-10} \\
		0 & 0 & 0
	\end{array} \right) ,
\end{align}
and 
%
the total lepton asymmetry is found to be $\eta^{\Delta L} = \sum_{l\alpha} \eta^{\Delta L}_{l\alpha} = 2.47 \times 10^{-8}$. This is well consistent with the observed value of the baryon asymmetry $\eta^{\Delta B} = \big(6.105^{+0.086}_{-0.081}\big)\times 10^{-10}$~\cite{Planck:2015xua}, after taking into account the sphaleron conversion rate and the entropy dilution factors.


\subsection{Lower bound of $m_{W_R}$}
Here we present our procedure to derive a lower bound of $m_{W_R}$ which is compatible with successful leptogenesis in a LRSM with all RH neutrinos being quasi-degenerate in mass. The total lepton asymmetry is given by Eq.~(\ref{eq:etaL}), which can be rewritten using Eq.~\eqref{wash} as
\begin{align}
	\eta^{\Delta L} (z) &= \frac{3}{4 \zeta (3)} z^2 K_1 (z) \sum_{l, \alpha} \frac{1}{W_l (z)}\varepsilon_{l \alpha}  \frac{\widetilde{D}_\alpha (z)}{D_\alpha (z) + S_\alpha (z)}   \; .
\label{eq:etaL1}
\end{align}
For the purpose of deriving a lower bound on $m_{W_R}$, we assume that the dominant washout effect comes from the $W_R$-mediated scattering processes, i.e., $S_\alpha \approx \sum_k S_{R \alpha} = 3 S_{R \alpha}$ where $S_{R \alpha} \equiv z\widetilde{\gamma}^{S_R}_{k \alpha}  / (n^\gamma H_N)$. This follows from the fact that  gauge interactions are flavor-blind,  and hence, $\widetilde{\gamma}^{S_R}_{k \alpha}$ has no dependence on lepton flavors. Similarly, when all the RH neutrino masses are quasi-degenerate, we can write $S_R \equiv \sum_\alpha S_{R \alpha} \approx 3 S_{R \alpha}$. We also assume that $D = \sum_\alpha D_\alpha \approx 3 D_\alpha$ for simplicity. Note that $S_R$ is not summed over lepton flavors while $D$ is. We further define $\varepsilon^Y_{l \alpha}$ and $B^Y_{l \alpha}$ as the $\CP$ asymmetry and branching ratio, respectively, without the 3-body decay width included in the denominator of Eqs.~\eqref{epsilon} and \eqref{BR}. For simplicity, we further assume that the branching ratios of the decay process $N_\alpha \to L_l \phi$ are the same for all the lepton flavors. With all the above-mentioned reasonable assumptions, we can approximate the total lepton asymmetry in Eq.~\eqref{eq:etaL1} at the critical temperature  as
\begin{align}
	|\eta^{\Delta L} (z_c)| &\approx \frac{3}{4 \zeta (3)} z_c^2 K_1 (z_c) \sum_l \frac{1}{\sum_\alpha [B_{l \alpha} D_\alpha (z_c) + S_{R \alpha} (z_c)]} \left| \sum_\alpha \varepsilon_{l \alpha} \frac{\widetilde{D}_\alpha (z_c)}{D_\alpha (z_c) + 3 S_{R \alpha} (z_c)} \right| \\
	&\approx \frac{9}{4 \zeta (3)} \frac{z_c^2 K_1 (z_c)}{S_R (z_c)} \frac{r_s r_d^2}{(3 + r_s r_d) (3 + r_s)} \varepsilon^Y_\text{tot}
	\label{eq:etaL2}
\end{align}
where the ratios $r_s \equiv D (z_c) / S_R (z_c)$ and $r_d \equiv \widetilde{D} (z_c) / D (z_c)$ parametrize the relative washout strengths of the 2-body decay, $W_R$-mediated decay and $W_R$-mediated scattering processes. The values of $r_s$ and $r_d$ depend on $m_N$, $m_{W_R}$, and the Yukawa coupling $h$. Assuming a specific value of $h$ and calculating the 2-body decay width simply as $\Gamma (N_\alpha \to L_l \phi) = \Gamma (N_\alpha \to L_l^c \phi^c) = h^2 m_N / 16 \pi$, we can evaluate these parameters as functions of $m_N$ and $m_{W_R}$.

\begin{figure}[t]
	\centering
		\includegraphics[width = 10 cm]{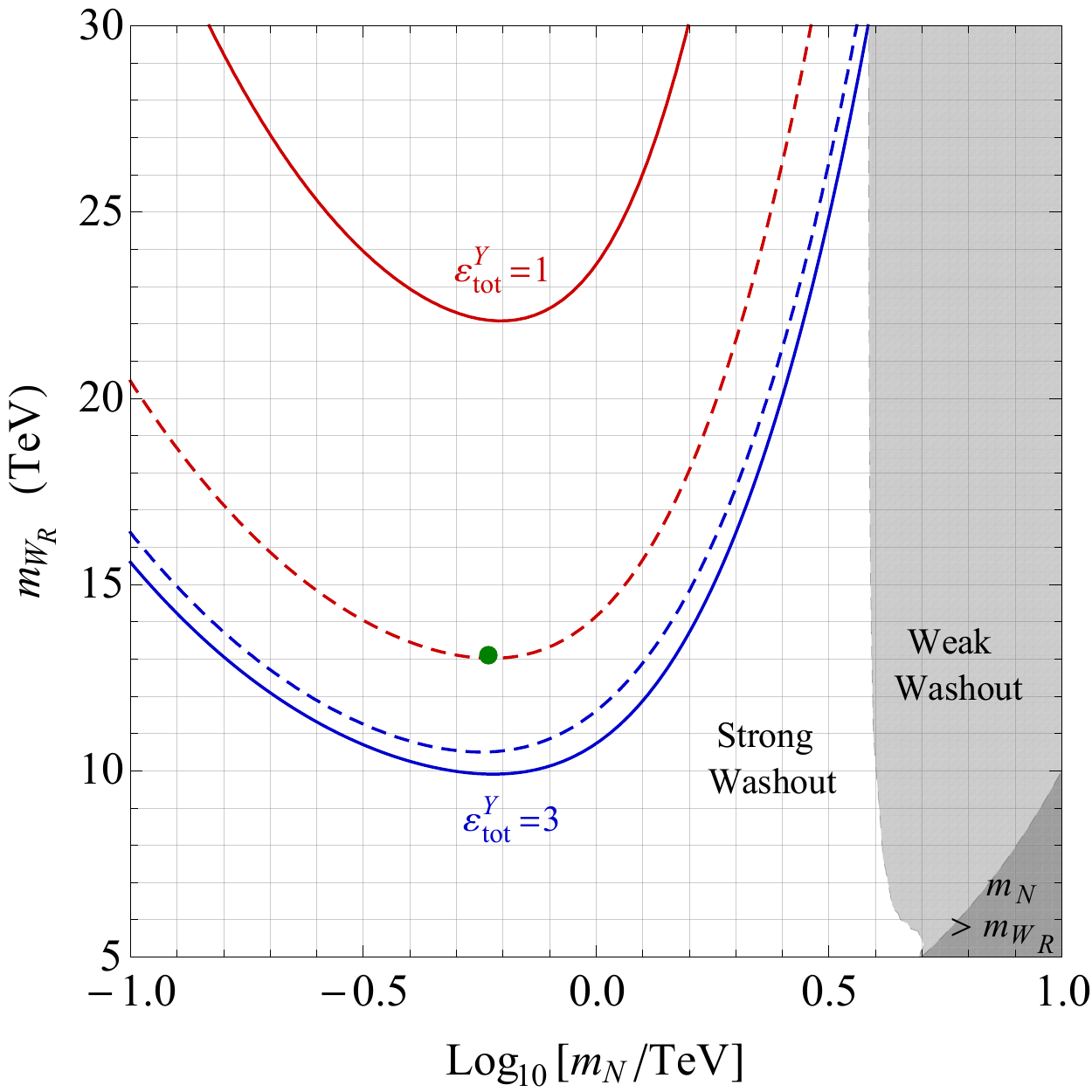} 
	\caption{Contour plots of $|\eta^{\Delta L} (z_c)| = 2.47 \times 10^{-8}$ for $h = 10^{-3.8}$ (dashed lines) and $h = 10^{-3.5}$ (solid lines) with $\varepsilon^Y_{\rm tot}=1$ (red lines) and  $\varepsilon^Y_{\rm tot}=3$ (blue lines). The green dot corresponds to the example fit value presented in Section~\ref{sec3}.}
	\label{fig:etaLCon}
\end{figure}

Figure~\ref{fig:etaLCon} shows the contour plots of $|\eta^{\Delta L} (z_c)| = 2.47 \times 10^{-8}$ for two different Yukawa couplings $h = 10^{-3.8}$ and $h = 10^{-3.5}$. The red curves correspond to $\varepsilon^Y_\text{tot} = 1$ which is the total $\CP$ asymmetry in the example fit given in Section~\ref{sec3}. Any region outside the red curve is incompatible with successful leptogenesis under the assumptions we introduced to obtain it, i.e. $m_{N1} \approx m_{N2} \approx m_{N3}$, $B^Y_{1 \alpha} \approx B^Y_{2 \alpha} \approx B^Y_{3 \alpha} \approx 1/3$, $\varepsilon^Y_\text{tot} = 1$ for the two specific values of $h$. For $h = 10^{-3.8}$,  we find that the lowest value of $m_{W_R}$ allowed for  $\varepsilon^Y_\text{tot} = 1$ is 13 TeV at around $m_N$ = 580 TeV. Note that the example fit we have presented in the previous section has $m_{W_R} = 13.1$ TeV, $m_N = 585$ GeV, and the resummed Yukawa coulings of order around $\sim 10^{-3.8}$. Thus, our  example fit, shown as a green dot in Figure~\ref{fig:etaLCon}, is very close to the minimum value of $m_{W_R}$ obtained here in a simplified manner, which justifies the validity of this approach. In other words, we have derived the lower bound in two different ways: (i) in Section~\ref{sec3} by carefully scanning the parameter space in order to find an explicit fit with the lowest possible $m_{W_R}$ compatible with leptogenesis, and (ii) using the approximate expression in Eq.~(\ref{eq:etaL2}). The remarkable  agreement between these two approaches shows that the simplified expression in Eq.~(\ref{eq:etaL2}) is very effective in finding the region in parameter space compatible with leptogenesis and predicting the lower bound of $m_{W_R}$ as well as the position in the parameter space where it exists. If we use the same expression and take the maximal $\CP$ asymmetry allowed in principle, i.e. $\varepsilon^Y_\text{tot} \equiv \sum_{l,\alpha} \varepsilon^Y_{l\alpha} = 3$, then we obtain the blue curves Figure~\ref{fig:etaLCon}. With this maximal $\CP$ asymmetry, we have found that the Yukawa coupling $h = 10^{-3.5}$ gives the lower bound of $m_{W_R}$ = 9.9 TeV at $m_N$ = 630 GeV.

\begin{figure}[t]
	\centering
		\includegraphics[width = 15 cm]{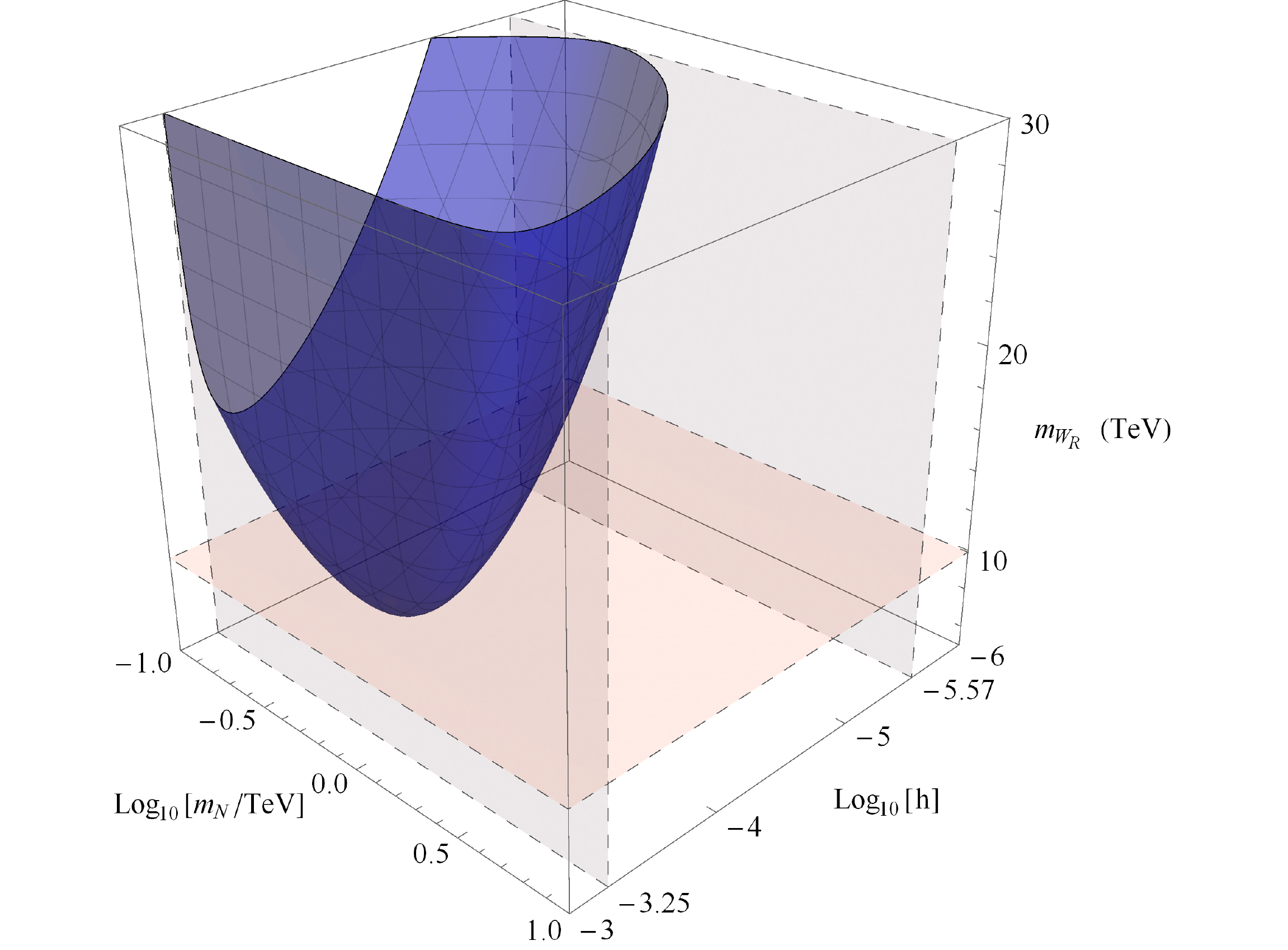} 
	\caption{The allowed region of parameter space (shaded region) yielding successful leptogenesis in our L-R seesaw model. The vertical gray surfaces show the bound on Yukawa couplings, while the horizontal pink surface shows the lower bound on $m_{W_R}$.}
	\label{fig:3D}
\end{figure}

The Yukawa couplings cannot be increased arbitrarily without spoiling the lepton asymmetry, since not only the source term due to the two-body decay of the RH neutrinos, but also the washout effects due to inverse decay and $\Delta L=2$ scattering increase with the Yukawa couplings.  Similarly, for very small values of the Yukawa couplings, the branching fraction of the two-body decay mode becomes comparable or smaller than the three-body decay mode due to $W_R$ interactions, and therefore, the washout effect again increases. Thus, successful leptogenesis works only in a range of the Yukawa coupling parameter space. This is shown in the three-dimensional plot given in Figure~\ref{fig:3D}, where we see that leptogenesis constraints in our model require the Yukawa coupling to be $10^{-5.6}\leq h \leq 10^{-3.2}$ for $m_{W_R}\leq 30$ TeV. The robustness of the lower bound on $m_{W_R}$ obtained in Figure~\ref{fig:etaLCon} can also be verified from Figure~\ref{fig:3D}.  

\section{Summary}\label{sec5}
In this proceedings, we address two issues related to seesaw models for neutrino masses. The first one deals with whether the TeV scale can be naturally in the TeV range without fine-tuning of parameters. We present a natural TeV scale left-right model which achieves this goal and therefore provides a counterexample to the common lore that either the seesaw scale must be superheavy or the active-sterile neutrino mixing must be tiny in a UV-complete seesaw model. The second issue addresses the important question: Whether in such low scale models, one can have successful leptogenesis, and if so, what constraints are implied by this on the mass of the RH gauge boson. In an explicit TeV-scale LR model, with an explicit fermion mass fit, we find the lower bound to be 13.1 TeV and for generic models in this class of L-R seesaw with enhanced neutrino Yukawa couplings compared to the canonical seesaw case and with maximal possible $\CP$ asymmetry for each flavor, this bounds becomes 9.9 TeV.

\section*{Acknowledgement} The work of P. S. B. D. is supported by the Lancaster-Manchester-Sheffield Consortium for Fundamental Physics under STFC grant ST/L000520/1. The work of  C. H. L. and R. N. M. is supported by the NSF grant No. PHY-1315155.

\appendix 
\section {Collision Rates} \label{app}
The various collision rates for decay and scattering processes in Eqs.~(\ref{be1}) and (\ref{be2}) are given below:
\begin{align}
\gamma^D_{l\alpha} \ & = \ \gamma^{N_\alpha}_{L_l\phi_l} + \gamma^{N_\alpha}_{l_Rq\bar{q}'},
\label{decay} \\
\widetilde{\gamma}^D_{l\alpha} \ & = \ \gamma^{N_\alpha}_{L_l\phi_l},\\
\gamma^{S_L}_{l\alpha} \ & = \ \gamma^{N_\alpha L_l}_{Qu^c}+\gamma^{N_\alpha u^c}_{L_l Q^c} + \gamma^{N_\alpha Q}_{L_l u} + \gamma^{N_\alpha L_l}_{\phi^\dag V_\mu} + \gamma^{N_\alpha V_\mu}_{L_l \phi} + \gamma^{N_\alpha \phi^\dag}_{L_l V_\mu}, \\
\widetilde{\gamma}^{S_L}_{l\alpha} \ & = \ \frac{\eta^N_\alpha}{\eta^N_{\rm eq}}\gamma^{N_\alpha L_l}_{Qu^c}+\gamma^{N_\alpha u^c}_{L_l Q^c} + \gamma^{N_\alpha Q}_{L_l u} + \frac{\eta^N_\alpha}{\eta^N_{\rm eq}}\gamma^{N_\alpha L_l}_{\phi^\dag V_\mu} + \gamma^{N_\alpha V_\mu}_{L_l \phi} + \gamma^{N_\alpha \phi^\dag}_{L_l V_\mu}, \\
\gamma^{S_R}_{l\alpha} \ & = \ 
\gamma^{N_\alpha l_R}_{\bar{u}_R d_R} + \gamma^{N_\alpha \bar{u}_R}_{l_R \bar{d}_R} + \gamma^{N_\alpha d_R}_{l_R u_R},\\
 \widetilde{\gamma}^{S_R}_{l\alpha} \ & = \ 
\frac{\eta^N_\alpha}{\eta^N_{\rm eq}} \gamma^{N_\alpha l_R}_{\bar{u}_R d_R} + \gamma^{N_\alpha \bar{u}_R}_{l_R \bar{d}_R} + \gamma^{N_\alpha d_R}_{l_R u_R},\\
\gamma^{(\Delta L=2)}_{lk} \ & = \ \gamma'^{L_l\phi_l}_{L_k^c\phi_k^\dag}+\gamma^{L_l L_k}_{\phi_l^\dag \phi_k^\dag}, \\
\gamma^{(\Delta L=0)}_{lk} \ & = \ \gamma'^{L_l\phi_l}_{L_k\phi_k}+\gamma^{L_l\phi_l^\dag}_{L_k\phi_k^\dag}+\gamma^{L_l L_k^c}_{\phi_l \phi_k^\dag} \; .
\end{align}
The scattering terms involving two heavy neutrinos in the initial state, e.g. induced by a $t$-channel $W_R$ or $e_R$, and by an $s$-channel $Z_R$, are not included here since their rates are doubly Boltzmann-suppressed and numerically much smaller than the scattering rates given above~\cite{hambye, Blanchet:2009bu}. The decay rates are explicitly given by
\begin{align}
\gamma^{N_\alpha}_{L_l\phi} \ & = \ \frac{m_{N_\alpha}^3}{\pi^2 z}K_1(z)\left[\Gamma(N_\alpha\to L_l \phi)+\Gamma(N_\alpha\to L_l^c\phi^\dag)\right], \\
\gamma^{N_\alpha}_{l_Rq\bar{q}'} \ & = \ \frac{m_{N_\alpha}^3}{\pi^2 z} K_1(z)\left[\Gamma(N_\alpha\to l_R q_R\bar{q}'_R)+\Gamma(N_\alpha\to \bar{l}_R\bar{q}_R q'_R)\right] \; .
\end{align}
The various collision terms for the $2\leftrightarrow 2$ scattering processes $XY\leftrightarrow AB$ are defined as
\begin{eqnarray}
\gamma^{XY}_{AB} \ = \ \frac{m_{N_1}^4}{64\pi^4 z}\int _{x_{\rm thr}}^\infty dx\sqrt{x} K_1(z\sqrt{x})\hat{\sigma}^{XY}_{AB}(x), \label{scat}
\end{eqnarray}
where $x=s/m_{N_1}^2$ with the kinematic threshold value $x_{\rm thr} = {\rm max}[(m_X+m_Y)^2,(m_A+m_B)^2]/m_{N_1}^2$, and $\hat{\sigma}^{XY}_{AB}(x)$ are the relevant reduced cross sections, whose explicit expressions can be found in~\cite{DLM2}. The reduced cross section of the additional process $W^-_R l^+_R\to W^+_R l^-_R$ mediated by $\Delta_R^{++}$ is given by 
\begin{align}
	\widehat{\sigma}_{W_R l_R}^{W_R l_R} (s) \ = \ \frac{g_R^4 m_{N_1}^2}{8 \pi s} \left[ \log{\left( \frac{(s - m_{W_R}^2)^2 + m_{\Delta_R^{++}}^2 s}{m_{\Delta_R^{++}}^2 s} \right)} + \frac{m_{\Delta_R^{++}}^2 s}{(s - m_{W_R}^2)^2 + m_{\Delta_R^{++}}^2 s} - 1 \right] .
\end{align}


\end{document}